\documentclass[draftclsnofoot,conference]{IEEEtran}

\usepackage{graphicx}
\usepackage{shorttoc}  
\usepackage{fancyhdr}

\usepackage{url}

\begin{document}
\pagestyle{plain}
\begin{onecolumn}
\thispagestyle{empty}
\title{Automation of PRL's\\ Astronomical  Optical Polarimeter with a GNU/Linux based distributed control system}

\author{\authorblockN{SHASHIKIRAN GANESH\authorrefmark{1},
U. C. JOSHI, K. S. BALIYAN, S. N. MATHUR, P. S. PATWAL and R. R. SHAH}
\authorblockA{Physical Research Laboratory\\Astronomy \& Astrophysics,
Ahmedabad, INDIA 380 009\\
\authorrefmark{1}Email: {\em shashi@prl.res.in}\\
}
}

\thispagestyle{empty}
\maketitle

\vfill
\begin{figure}[h]
\centerline{\includegraphics[width=2cm]{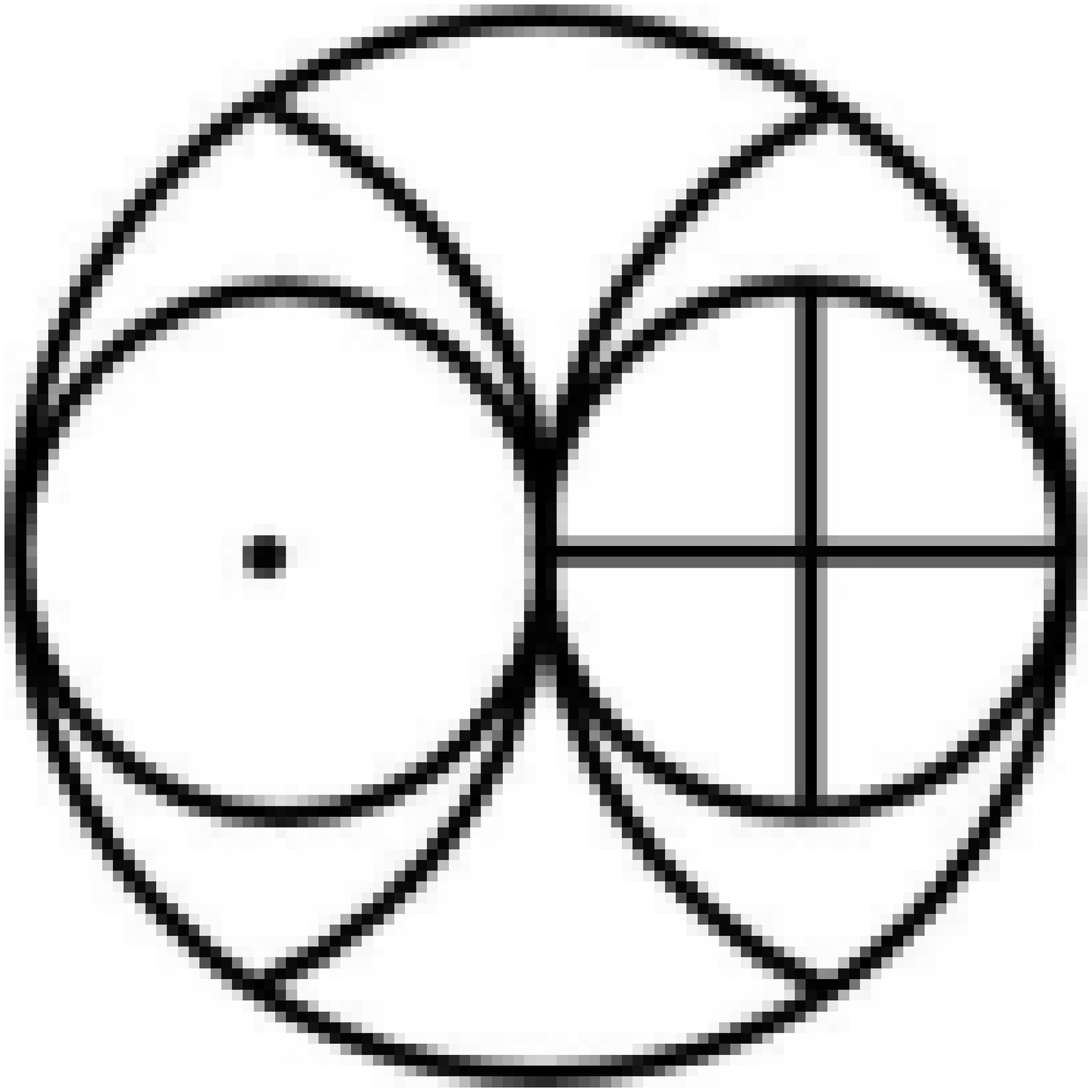}}
\end{figure}
\begin{center}
{\large Astronomy \& Astrophysics Division}\\[0.5cm]
{\Huge Physical Research Laboratory}\\[0.8cm]
{\large Ahmedabad, INDIA 380 009}\\

\end{center}
\thispagestyle{empty}

\newpage
\thispagestyle{empty}

\noindent
This document was created by the authors using \LaTeX ~with a style file based on the IEEE style file 
\noindent
and {\em fancyhdr, graphicx} \LaTeX ~packages 

\noindent
Original version submitted September 2008, accepted March 2009\\

\vfill

\begin{center}
\url{http://www.prl.res.in/~shashi/}\\ 
\url{http://www.prl.res.in/~library/}\\
PRL Technical Report \# PRL-TN-2008-93
\end{center}

\vfill

\noindent
This work is licensed under the Creative Commons Attribution 3.0 Unported License. To view a copy of this license, visit http://creativecommons.org/licenses/by/3.0/ or send a letter to Creative Commons, 171 Second Street, Suite 300, San Francisco, California, 94105, USA.\\

\newpage
\pagestyle{empty}
\shorttoc {Short contents} {1}
\newpage
\label{toc}
\tableofcontents
\newpage

\end{onecolumn}
\begin{twocolumn}

\setcounter{page}{1}
\pagestyle{fancy}

\begin{abstract}

PRL's Optical Polarimeter has been used on various telescopes in India since its development in-house in the mid 1980s.  To make the instrument more efficient and effective we have designed the acquisition and control system and written the software to run on the GNU/Linux Operating System.  CCD cameras have been used, in place of eyepieces, which allow to observe fainter sources with smaller apertures.  The use of smaller apertures provides dramatic gains in the signal-to-noise ratio.   The polarimeter is now fully automated resulting in increased efficiency.  With the advantage of networking being built-in at the operating system level in GNU/Linux, this instrument can now be controlled from anywhere on the PRL local area network which means that the observer can be stationed in Ahmedabad / Thaltej as well or via ssh anywhere on the internet.   The current report provides an overview of the system as implemented.  

\end{abstract}

\section{Introduction}

The Optical Polarimeter (see Fig. \ref{opt_scope}) has been in use as one of the backend instruments (Deshpande et al., 1985, Joshi et al., 1987) at the 1.2m telescope operated by the Astronomy \& Astrophysics Division of the Physical Research Laboratory (PRL) at Gurushikhar near Mt Abu. 

This instrument enables the study of polarisation at optical wavelengths of a wide variety of astronomical subjects ranging from comets and stars to blazars.  To minimize the error due to the background sky light, one should use the smallest apertures possible (say 6 to 10 arc sec), however in the case of visual centring of the source this is not always possible due to the very small contrast between the typical sources of interest (such as blazars / quasars) and the sky.     In the absence of on-axis guiding, it was not possible to make long integrations to improve the signal-to-noise ratio.  
\begin{figure}[h]  
\centering
\includegraphics[width=0.8\columnwidth]{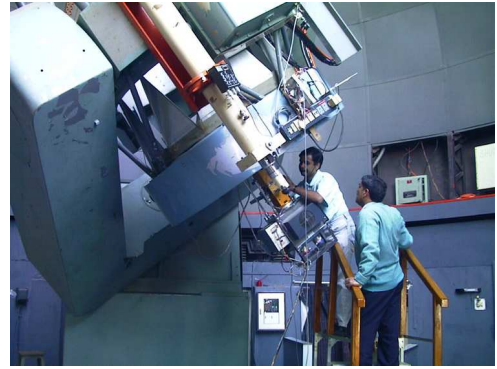}
\caption[Optical polarimeter on 1.2m Telescope (1997)]{\label{opt_scope}
The optical polarimeter mounted on the Cassegrain focus of the 1.2m Mt Abu telescope (circa 1997).}
\end{figure}

In order to improve the efficiency of the instrument and to address the above shortcomings  we have completely overhauled and rebuilt the acquisition system and added new subsystems.   To reduce human interaction and thus human error we have used CCD cameras, in place of eyepieces, which enable to look at the location of star vis-a-vis the edge of the aperture on a monitor.   The instrument has been fully automated using GNU/Linux with an RTAI (Real Time Application Interface) enabled kernel running on a PC/104 based embedded CPU board\footnote{PC/104 format is a compact ($96 \times 90 $mm$^{2}$ size) board with the ISA bus' 104 pins arranged in four rows in a condensed format.  With a  pin and socket connection, the PC/104 systems are self-stackable and are extremely rugged when compared to normal ISA bus based motherboards}.  Onyx PC/104 counter/timer and digital I/O boards were utilized to record the counts coming from photo-multipliers in photon counting mode.  One PC/104 board has been developed in house for rotating a half-wave plate to generate fast modulation of incoming light beam.  
Other mechanical operations (such as changing of optical filters, apertures etc) have been achieved through the use of stepper motors driven by Atmel microcontrollers.  Another PC/104 CPU board controls a USB interfaced CCD camera to provide a view of the observing aperture and the field being observed by the instrument.  An additional independent embedded computer controls the auto-guider CCD.   With the use of GNU/Linux and in-house developed control software (both kernel device driver module as well as user space) the instrument can be operated from anywhere on the local area network (LAN).   Since the automation, the instrument has been used extensively from the fully enclosed telescope control room adjacent to the telescope floor(dome) and also remotely from PRL campus at Ahmedabad.   In principle, the operator/observer can be stationed anywhere on the PRL computer network including Ahmedabad / Thaltej.  

In this report we provide a brief overview of the techniques employed to upgrade the instrument.  While the concepts discussed here have been implemented for an astronomical instrument, they are general enough to be applicable to other experimental sciences wherever remote control is desired.  Further reports in this series will elaborate on the various aspects in much greater depth.  This is done with a view to make the reports as modular and self-contained as possible so that information of interest is easy to locate for developing other instruments in future.

\section{Principle of operation}
\label{principle}

\begin{figure} 
\centering
\includegraphics[width=\columnwidth]{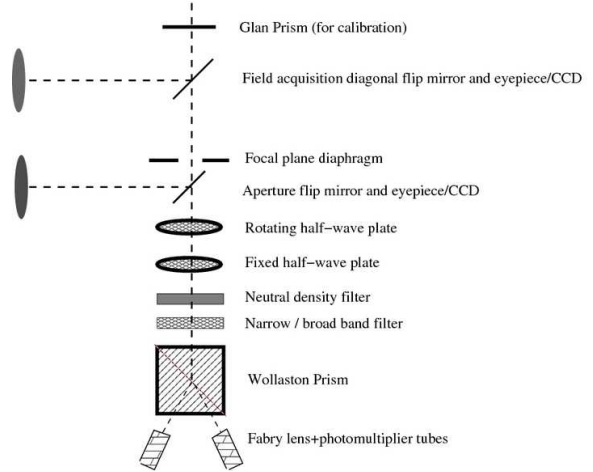}
\caption{\label{optical_layout}
Optical schematic of the polarimeter.  The narrow and broadband filters (typically 80\AA and 1000\AA bandwidth respectively) are the only components which need to be changed based on science requirement.  The half-wave plates are good from 3500 to over 10000\AA.  The neutral density filter reduces the incoming starlight by 2.5 magnitudes.}
\end{figure}
The principle of operation of the instrument is described by Frecker \& Serkowski (1976). The basic idea is to measure the optical polarisation by the use of an analyser, with photo-multiplier tubes (PMTs) recording the counts in photon counting mode.  The light path through the instrument is shown in the schematic layout in Fig. \ref{optical_layout}.  In order to minimise the influence of varying atmospheric conditions, fast modulation of the incoming light beam is used. For this purpose a half-wave plate is rotated by a stepper motor at  5 or 10 rotations per second with 96 steps per rotation resulting in sampling time of 1 or 2msec per step.   The modulated beam is then split into ordinary and extra-ordinary polarised components by a Wollaston prism and the respective counts registered by two independent photon counting photo-multiplier tubes.   Due to the modulation, the recorded counts exhibit a sine wave pattern in 24 steps.   Since this includes contribution from the sky, an equally large, vacant area of the sky adjacent to the source is observed.  These counts are subtracted from the counts recorded for the object of interest.   A function of the form 
\begin{equation}
I_{j}~=~{{1}\over{2}}{\Big\{I_{0}~\pm~Q~\cos 4\theta_{j}~\pm~U~\sin 4\theta_{j}\Big\}}
\end{equation}
is fitted to the counts $I_{j}$ recorded at different positions of the half-wave plate (angle $\theta_{j}$) and the Stokes parameters describing linearly polarised light $I_{0}$, $Q$ and $U$ are obtained.   From these the degree, $p$, and position angle, $\Theta$, of polarisation are readily obtained using the simple formulae:
\begin{equation}
p~=~\sqrt{Q^2+U^2}~~~\mathrm{and}~~~\Theta={{1}\over{2}}\tan^{-1}\bigg\{{{U}\over{Q}}\bigg\}
\end{equation}

\section{Hardware}
\label{hware}

A distributed embedded control system (see block diagram in Fig. \ref{schematic}) has been developed as described in the following sections.  The eyepieces have been replaced by CCD cameras from Starlight-Xpress which have proved to be extremely efficient in detection of the source and the edge of the aperture being used.  With 16 bit data contrast levels even very faint sources can be observed with ease now and they can be accurately centred even in the smallest of apertures.  For changing the apertures we have implemented a stepper motor driven rack and pinion coupled mechanical extension to the existing aperture slide.  A similar mechanism has been implemented for changing the optical filters.  A third motor unit allows to pull in and out the mirror which directs light to the CCD or to the photo-multiplier tubes.  

All of the power supplies, support electronics, computer boards are contained completely in an {\em Embedded Control System} box.  This makes the instrument a very efficient self-contained unit.  The entire instrument is mounted as a single unit on the Cassegrain focus of the telescope and only three cables need to be connected to the instrument :  A.C. mains power supply input, ethernet connectivity cable and finally a cable to interface the telescope guiding with the instrument.  

\begin{figure} 
\centering
\includegraphics[width=0.9\columnwidth]{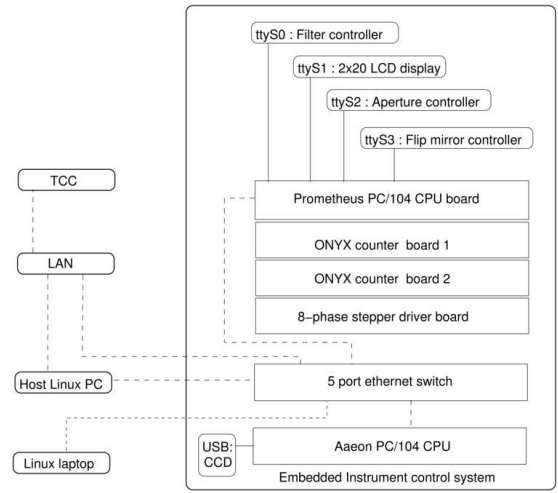}
\caption[Schematic of control and acquisition system]{\label{schematic}
Schematic of control and acquisition system of the polarimeter.  Dashed lines indicate ethernet connectivity between subsystems.  TCC is Telescope Control Computer.  The observer's computer is shown as a Linux laptop; this can be located anywhere on the local area network (LAN).}
\end{figure}

\subsection{PC-104 control system}

\begin{figure} 
\centering
\includegraphics[width=0.8\columnwidth]{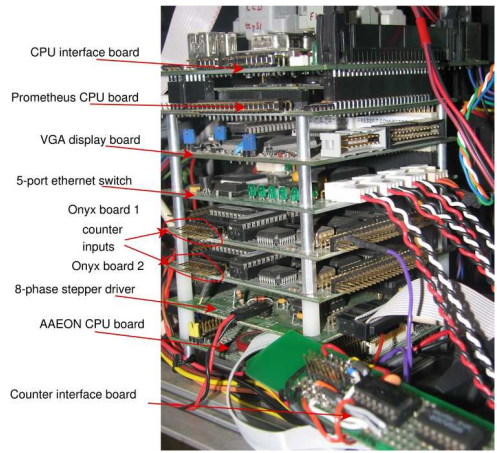}
\caption[The inner view of the embedded distributed control system showing the PC/104 stack]{\label{pc104stack}
The inner view of the embedded control system showing the PC/104 stack.   The counter interface board has been removed from it's usual position across the ONYX boards to show the PC/104 stack clearly.}
\end{figure}

The {\em embedded control system} mounted as a part of the instrument  consists of two embedded PC/104 CPU boards.  PC/104 specification is a compact ($90 \times 96$ mm$^2$) size bus based system.  The Prometheus PC/104 CPU board (manufactured by M/s Diamond Systems) controls the data acquisition process and distributes jobs to the other subsystems.  It is a Zfx86 CPU (equivalent to a 100MHz 486-DX2) board with 32MB RAM, 10/100 Mbps ethernet, 4 serial and one IDE port apart from other peripherals.  A 32MB solid-state IDE flash disk is connected to the IDE interface. This 32MB disk is sufficient for the entire operating system and control software (as discussed in the next section).     This is a self-stackable rugged system.  The ruggedness comes from the 104 pins of the PC/104 bus which are arranged in four rows on one side of the board.  The CPU board has both male and female bus connectors and other peripheral cards can be stacked on top of / below or on both sides of the CPU board.  The boards are supported on each corner by threaded PC/104 stand-off supports or spacers.   Thus the entire stack is electrically as well as mechanically ruggedly supported.  One needs special board separator or extractor tools to separate the PC/104 boards without damaging the bus pins.  In the stack that we have implemented (see Fig. \ref{pc104stack}) there are two CPU boards in the same physical stack but with the bus connections being independent.  The Prometheus CPU board is connected to several PC/104 peripheral boards - a VGA display board (used only for debugging purpose), a 5 port 100 Mbps ethernet switch, two ONYX digital I/O and counter/timer boards and an in house developed 8-phase stepper driver board (see next section).   The other PC/104 CPU board in this stack controls the CCD Camera connected via USB interface.  Since the data storage devices are flash based (i.e. semiconductor based), the reliability is orders of magnitude better than the earlier hard disk based systems.  This advantage arises from the lack of moving components in flash based storage devices.

\subsubsection{Stepper motor driver board}

An 8 bit PC/104 card (schematic shown in Fig. \ref{sch8phase})  was developed in-house for driving the legacy 8-phase stepper motor for the rotating half-wave plate.  This motor + gearbox coupling to the rotating half-wave plate has been working very smoothly for a very long time with occasional requirement to replace (once in 8-10 years) the precision carbon bearings of the rotating half-wave plate. This card was designed and built as a double layer PCB and uses an 8254 timer chip.   A 74LS164 8 bit-serial-to-parallel shift register is used for providing the 8 phase timing waveforms to the stepper motor via a ULN2803 driver IC.  The 8254 timer chip provides the clock pulses and is also a source of hardware interrupts to the Prometheus CPU every 2 msec  corresponding to the duration between each step of the stepper motor / half-wave plate.   For sensing a reference point in the rotation of the half wave plate the following arrangement has been made.   The gear wheel coupled to the stepper motor has a tiny hole near the edge of the wheel (rest of it being a solid block).   Fixed on the top of the gearbox is  a light emitting diode (LED) and on the bottom side across where the hole passes in front of the LED is a light dependent resistor (LDR).  The output of the LDR is suitably amplified and shaped as a pulse and this digital pulse is monitored by a digital I/O bit of one of the Onyx interface boards discussed below.  

\subsubsection{Onyx counter boards}

Two Onyx PC/104 counter boards were obtained commercially from the same vendor as the Prometheus board.  The Onyx counter and digital I/O board provides 16 bit counter timer functionality with the use of an 8254 timer chip.  For each of the two PMTs, we used two counters of the 8254 on one counter board.  Using suitable gating inputs derived from the interrupt pulse, described above, each of the two counters per PMT input is alternately enabled and disabled for counting in binary down counting mode.  When one counter is being read and reset, the photon counts are being recorded by the other counter.  We also tried the 9513 timer chip (which has 5 16bit counter/timers on a single chip) but this was not as successful at recording the PMT output pulses as the 8254 chip.   The 8254 (with 3 16-bit counter/timers) is able to record pulses as narrow as a few nano sec, while the 9513 chip requires that the pulse width be much larger (typically a few 100 nano sec). 

\subsection{CCD cameras}

Two CCD cameras have been used in place of the eyepieces shown in Fig. \ref{optical_layout}.  

\subsubsection{Starlight Xpress SXV-H9}

An SXV-H9 CCD camera from Starlight-Xpress is used to view the source and accurately centre it in the aperture being used.  This camera is a $1392 \times 1040$  pixel 16-bit thermoelectrically cooled CCD device with exceptionally compact driver electronics.  This is mounted in the location where the aperture eyepiece was previously located.   An other PC/104 CPU board (PCM-5330) sourced from M/s Aaeon Technology is used for controlling the CCD camera.  This CPU board is based around an STPC Atlas System-on-Chip (SoC : x86 equivalent) running at 133MHz.  It has 64MB on board RAM and 10/100 Mbps Ethernet, 4 serial and 2 USB ports along with an IDE and compact flash interface.  A 128MB solid-state compact flash disk is used for the operating system and control software with this board.  It is also connected to the main telescope controls via a 4 bit channel corresponding to North/South/East/West movements from the guider interface of the CCD camera.  

\subsubsection{Starlight Xpress SXVF-M25C}

The SXVF-M25C CCD Camera from Starlight-Xpress is a one-shot-colour CCD camera with a large field of view.   It has $3024 \times 2016 $ pixels in a Bayer matrix.  This CCD shares the same USB interfacing techniques as the SXV-H9 and is also connected with the telescope movements via the guider interface.     With it's large field of view it is used for the field acquisition and source identification.   This CCD is mounted in place of the field acquisition eyepiece (see Fig \ref{optical_layout}).    If need be, both CCDs can be interchanged.

\subsection{Micro-controller subsystems}

An AVR micro-controller PCB board has been developed in house.  We have designed and built this board around an Atmel AVR (ATMEGA8 or ATMEGA88) as the micro-controller and with a MAX 232 serial interfacing chip.  The PCB supports in circuit programming via a 5 pin connector (programming port).  This programming port can be connected to a parallel or USB port of a host PC with the appropriate cables.  This board has been used in all the modules discussed in the subsections below.   The basic PCB remains the same and minor changes are hand made by using the general purpose pin outs made available on the PCB.  The variations are mainly in the firmware for each application.  A separate technical note is in preparation which will present the hardware / software / firmware details of the AVR board based stepper motor controller.

\subsubsection{Stepper motor with discrete position encoding}

Three stepper motors are used in the movement of the various components i.e. the filter selection slide, the aperture selection slide and thirdly a sliding mirror to divert light from the rest of the instrument.  These are controlled by three identical stepper motor control cards and interface to the Prometheus PC-104 board via serial interfaces.  These stepper motor control cards are based on an Atmel Atmega 8 AVR microcontroller and were developed in house (schematic of the stepper motor control card is shown in Fig. \ref{avrsch}).  

\subsubsection{Monitoring temperature of cold chamber holding the PMTs}

One of the serial ports of the Prometheus motherboard interfaces to an Atmega 8 microcontroller which monitors other system parameters such as the voltage levels and temperatures (both ambient as well as temperature inside the cold box holding the PMTs).   The temperature monitoring is done with a DS18S20 one-wire sensor interfaced to one of the I/O pins of the microcontroller.  ADCs on the microcontroller are used for monitoring the various voltage levels (supply, control etc.) required for operating the PMTs.  This AVR board uses a copy of the same PCB as the stepper motor controller discussed in the previous section (without the ULN2003 driver IC being mounted).  

\subsubsection{LCD driver}

Yet another serial port on the Prometheus board is used to display status information on a character matrix LCD mounted on the embedded system.  This is again coupled via another AVR board which takes serial input and provides suitable glue logic to display it on the LCD.   

\subsection{Power supplies}

Two power supplies are used for powering the different subsystems in the embedded control system.  One compact 55 Watt SMPS with 5 and 12V output powers most of the electronics including the two PC/104 CPUs and the stepper motors.  Another SMPS provides 5V supply for powering the 8-phase stepper motor and 12V as input supply for the CCD cameras.    Compact high voltage power supply modules (total weight few hundred gm) from Electron Tubes have been used in place of the original bulky (several Kg) power supply.   These h.v. power supply modules require 24V input and their output can be monitored.     The PMTs are housed in a cold box where the temperatures are held at about 30 degree below ambient temperature.   The linear power supply being used for the thermoelectric cooling unit has been replaced with a high current SMPS power supply which weighs a fraction of the original supply and is also much reduced in terms of volume.  All these supplies which were earlier housed in individual chassis and mounted separately on the telescope or kept on the observing floor table are now made an integral and permanent part of the instrument and need not be disconnected for storage between observing runs. 

\section{Software}
\label{sware}
\subsection{Operating system}
\begin{figure}[h]  
\centering
\includegraphics[width=\columnwidth]{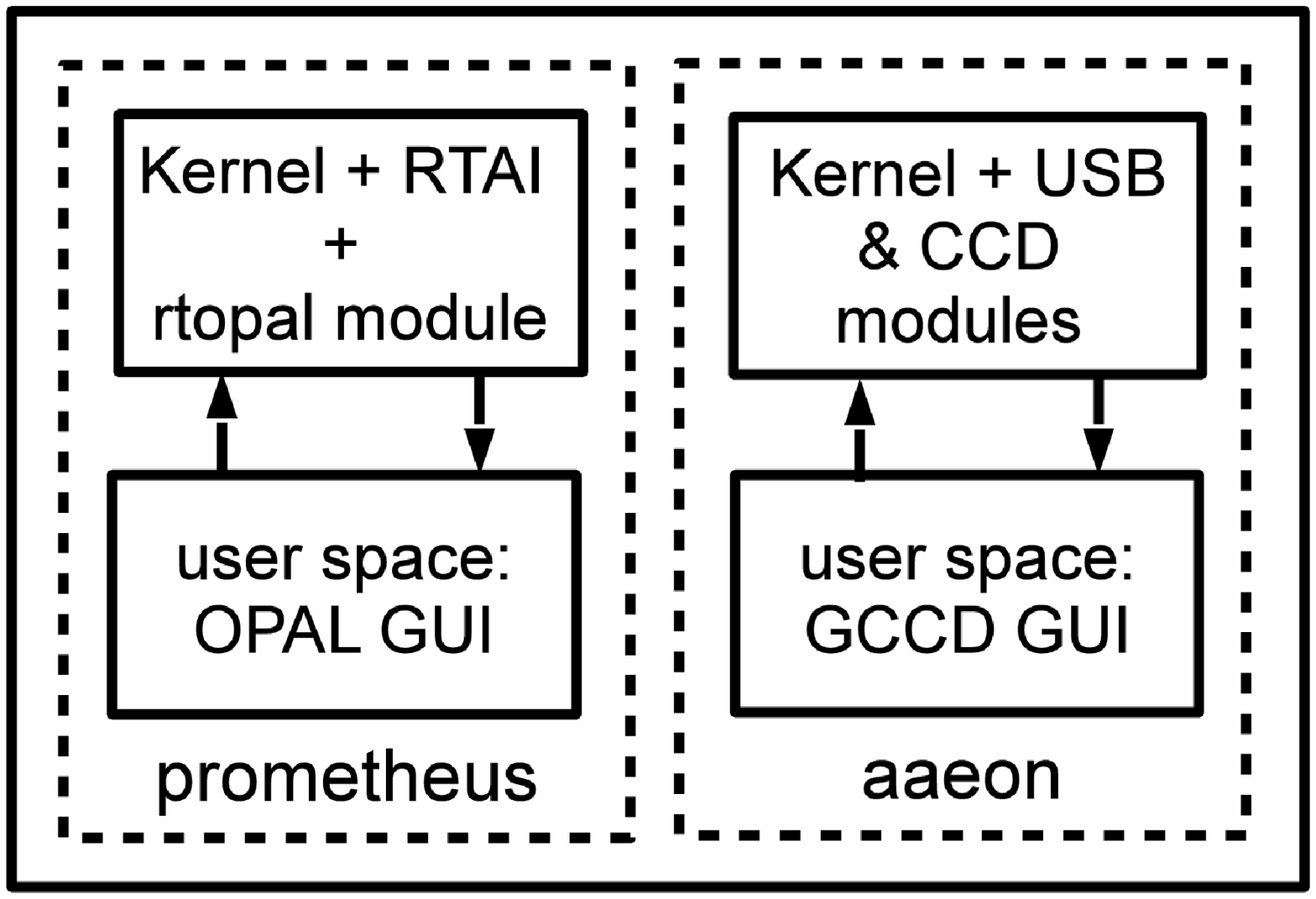}
\caption[Software block diagrams]{\label{softwareblocks}
Block diagram of the kernel and user space software on the Prometheus and Aaeon PC-104 linux systems. The different kernel and user modules are discussed in the text.}
\end{figure}

We make use of the GNU/Linux Operating System for the control of the instrument.   This is a unix like operating system available for a large variety of microprocessors.  It is easily scalable from 32 bit AVR microprocessors to high end clusters (super-computers).   We have been using this OS for the analysis of astronomical data from most of our instruments (CCD and NICMOS images etc.) and all analysis and developmental software are freely available for this OS under the GNU General Public License (GPL) or other similar open source licenses.   One of the biggest advantages of this operating system environment (compared to the single tasking MS-DOS) is that it is fully multi-tasking and network interfacing is fully built-in at the very basic level (Kernel) of the operating system.  Highly advanced graphical user interfaces are available and high level libraries (both general computation as well as scientific application related) are easily available with full documentation.   Virus related problems seen in other operating systems such as MS-DOS and MS-Windows are not present in GNU/Linux.  The scalability of the OS is such that a minimal system with networking support can be fit in to less than 2MB of disk space.    

The instrument is controlled by a dedicated control PC over LAN.  This control PC is usually kept in the control room at the observatory and serves as the operator console hosting the X-Window graphical interface.   It is a Pentium III running at 800MHz with 512MB RAM with standard Redhat 7.3 Linux distribution along with all required developmental tools/software.  The PC is connected via Ethernet cable to one of the ports of the 5-port Ethernet switch of the embedded system.  This control PC exports it's home partition as a network file system(NFS).   However, the instrument is completely independent of this PC and can be controlled from any system with a network reachable X-Window display.  In the event of the home partition not being available via NFS one can save the data on USB flash drives or other devices connected directly to the PC/104 stack.

The Prometheus board runs GNU/Linux with real time extensions (RTAI : Real Time Application Interface) added to a standard Linux kernel (version 2.4.19) from \url{www.kernel.org}.  The file system on the 32MB flash disk is based on white-dwarf Linux.  The base operating system requires only 16MB. Additional space is taken up by the GTK graphical interface libraries and the application software.  The data recorded by the  system is saved on the NFS (network file system) partition mounted as /home on the embedded Prometheus CPU board. 

In Fig. \ref{softwareblocks} we show the block diagrams of the software implementation on the two PC104 systems (Prometheus for the main polarimeter system and Aaeon for the CCD sub-systems).   As shown, both user and kernel space codes have been developed for this instrument and are described in the following sub-sections. 

\subsection{Kernel space drivers}
\subsubsection{Stepper driver board and Onyx counter boards}

The control software consists of both kernel space as well as user space code.  Kernel level codes (marked as {\em rtopal} in Fig.\ref{softwareblocks} initialise all the 3 PC-104 interface boards (2 Onyx boards and the 8-phase stepper driver board).    The integration or exposure starts after the starting position has been sensed by monitoring the status of one of the digital I/O bits connected to the LDR via a pulse shaping circuitry.  Thereafter the exposure continues until the specified time interval has been completed.  The job of reading out the counts in synchronisation with the interrupts received at each step of the half wave plate is  carried out in real time kernel module code written in C.  In order to remove the effects due to the jitter  in the interrupt response and the finite time it takes to record the counts from the 8254 counter, we use two counters per PMT as mentioned earlier.  During the first 2msec one of the counters is enabled and is down counting.  At the end of the 2msec the first counter is disabled by a suitable gating level and it is read out by the host processor and then reset.  At the same time an inverted gate is supplied to the second counter which starts counting down until the end of 2msec and so on.  The same process is followed for the second PMT+counter board combination at the same time. All other functions are disabled during the time the system is recording the counts from the celestial sources (which can be of typically few seconds to few minutes in duration).   By using hardware gating we have precisely equal intervals for each readout independent of any interrupt jitter that is always present in a multi-tasking OS (although that in itself is also minimised with the RTAI extensions).    The device driver software code is available from the authors.  

\subsubsection{CCD USB device driver} 

The USB device driver for the two CCD cameras is derived from the code originally written by David Schmenk.   The SXV-H9C CCD camera is directly supported by {\em ccd\_kernel} version 1.8.     In the case of the SXVF-M25C camera, the {\em ccd\_kernel} driver had to be modified to include appropriate device parameters and also to adapt the code to the special read-out mode of the CCD chip.  

\subsection{User space GUIs}

The instrument is controlled by Graphical User Interface (GUI) software which run on the PC/104 sub-systems with the display being provided by the local X-Window terminal of the observer.  This could typically be the observer's laptop or any desktop on the local-area-network.    Two graphical interfaces are launched.   One, called {\em OPAL}, is for controlling the basic instrument and acquiring and displaying the data.  The second one is {\em GCCD} for the control of the CCD cameras.

\subsubsection{GUI : OPAL controls}

Graphical interface software {\em OPAL} designed using the {\em GLADE} software runs in user space.  It is written in {\em C}, and uses only the {\em GTK} graphical libraries.  This complies with tight memory and execution time constraints.  The {\em OPAL} software also interfaces with the telescope control computer (TCC) over LAN, using network sockets, to record the telescope parameters (time, direction etc.) at the time of observation.  
The aperture and filter selection is menu-driven.  The callbacks from the respective menu option send command codes to the respective microcontroller boards via independent serial ports.  The status feedback from each microcontroller is displayed in the window and also recorded with the computed output of each observation.     Several {\em C} code and header files implement the details of the user interface and callbacks etc.   The compilation is via the standard GNU `{\em make}' mechanism.  A tar file containing the entire source code is available on request from the authors.  
\begin{figure}[h] 
\centering
\includegraphics[width=0.8\columnwidth]{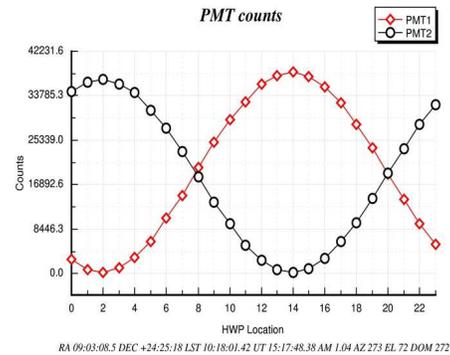}
\caption{\label{100percent}
100\% polarisation as observed with a Glan prism for star $69~\nu~ Cnc$}
\end{figure}

The individual observation records are saved incrementally to a text format file along with the telescope parameters.   The file name is derived from the  date of observation and is opened in append mode so that  all observations of a given night are contained in one file.    For each observation we also save the individual counts recorded at the 24 folded positions of the half-wave plate in a separate file.        Fig. \ref{100percent} shows a plot of test observation for 100\% polarisation.  This is a plot saved in postscript format by the {\em OPAL} GUI.  The test for 100\% polarisation is done by introducing a Glan prism in the light path and observing a bright star.   Typical measurements range from 97.5 to 99.5\%.  Compliance with observation of 100\% polarisation demonstrates the overall linearity of the system (from very low counts to a few million counts) in the polarisation measurements.  

\subsubsection{GUI  : CCD Controls}

The Aaeon PC/104 CPU board has more resources in terms of memory and operating system base space so we have installed a very stripped down version of Redhat 7.3 Linux distribution on the 128MB compact flash disk of this board.  The Starlight-Xpress SXV-H9 CCD is operated by a free software called {\em GCCD} written by David Schmenk.  The software uses the gnome library files and so has a little larger memory requirement than the {\em OPAL} GUI.  {\em GCCD} is available on the website listed in the resources below.

As already mentioned it is also possible to make small movements of the telescope to accurately centre the source in the aperture while monitoring the CCD view.  Thus the instrument is fully integrated with the telescope control system. 

\subsection{AVR firmware}

All the five Atmega 8 microcontrollers used in this instrument were programmed using a version of C (GNU-compiler collection - {\em gcc}) for the AVR again on a GNU/Linux PC with the appropriate developmental tools (compilers / libraries).   Useful programming tips and tools (including a bootable {\em live-cd} with compilers and other software tools for AVR programming are available on the PHOENIX project (Physics with Home-made Equipment \& Innovative Experiments) website of the Inter-University Accelerator Centre (IUAC)  and also on the Tuxgraphics.org websites.  A separate technical note is in preparation and details the software(firmware) and hardware aspects of the use of the AVR Atmega PCB board.    The PCB is general purpose enough to be usable as a microcontroller experimental and developmental board. 

\section{Summary}

We have designed and built the electronics and control system of the Optical Polarimeter.   It is controlled by software running on a GNU/Linux/RTAI platform and can be controlled from anywhere on the local area network.    Hardware and software including firmware for the micro-controllers were developed completely in-house.   Use of CCD cameras in place of the conventional eyepieces allows to observe very faint sources systematically and efficiently.      Precise centring of the source in the observing aperture is now possible routinely.  This has also allowed to use much smaller apertures (6 to 10 arc sec) than what was being used earlier (15 to 20 arc sec) for the observations.   With the smaller apertures, sky (background) contribution reduces and thus the noise due to the background also reduces.   This provides significant gain in the signal-to-noise (S/N) ratio and also enables to observe much fainter sources than was possible earlier.   Human error has been nearly completely taken out of the picture as far as the observational aspects are  concerned.   

\section{Acknowledgements}

The 8-phase stepper motor driver board was built as part of an M.Sc. project
carried out by students (Nirmit Dudhia and Prashant Raghuvanshi of Gujarat
University) while some of the AVR microcontroller codes were implemented by
Gagan Mallik, student of Nirma University as part of his B.E. project under our
supervision and guidance.   We acknowledge useful discussions with N.M. Vadher, 
A. B. Shah and C. R. Shah.  We are thankful to our colleagues  in the Astronomy
and Astrophysics Division and the staff members at Mt Abu Observatory for their
help and support.  We also acknowledge the help provided by PRL workshop.   This report was prepared with \LaTeX with a style file modified from the IEEE format.  This
work is supported by the Dept. of Space, Govt. of India.\\[0.5cm]

\section{References}
\begin{itemize}
\item Deshpande, M. R., Joshi, U. C.,  Kulshrestha, A., Banshidhar, Vadher,  N.
M., 1985,  An astronomical polarimeter, Bulletin of the Astronomical Society of 
India, v. 13, pp. 157-161.
\item Frecker, J. E., Serkowski, K., 1976, Linear polarimeter with rapid modulation, achromatic in the 0.3-1.1-micron range, Applied Optics, v. 15. pp. 605-606.
\item Joshi, U. C., Deshpande, M. R., Sen, A. K., Kulshrestha, A., 1987, Polarisation
investigations in four peculiar supergiants with high IR excess, Astronomy  \&
Astrophysics, v. 181.  pp. 31-33.\\[0.5cm]
\end{itemize}
\section{WWW Resources}
\begin{itemize}
\item Free Software Foundation website : \url{http://www.fsf.org/}
\item RTAI : \url{http://www.rtai.org/}
\item Linux kernel website : \url{http://www.kernel.org/}
\item White Dwarf Linux web pages : \url{http://www.blast.com/index.php?id=66}
\item {\em GLADE} user interface design software : \url{http://glade.gnome.org}
\item David Schmenk's {\em GCCD} software website : \url{http://schmenk.is-a-geek.com/} 
\item \url{http://www.iuac.res.in/~elab/phoenix/}
\item \url{http://tuxgraphics.org/electronics}
\item The source codes accompanying this report and further figures are available on the first author's webpages at \url{http://www.prl.res.in/~shashi/inst.html}
\end{itemize}

\begin{figure*}[hb] 
\centering
\includegraphics[width=0.8\textwidth]{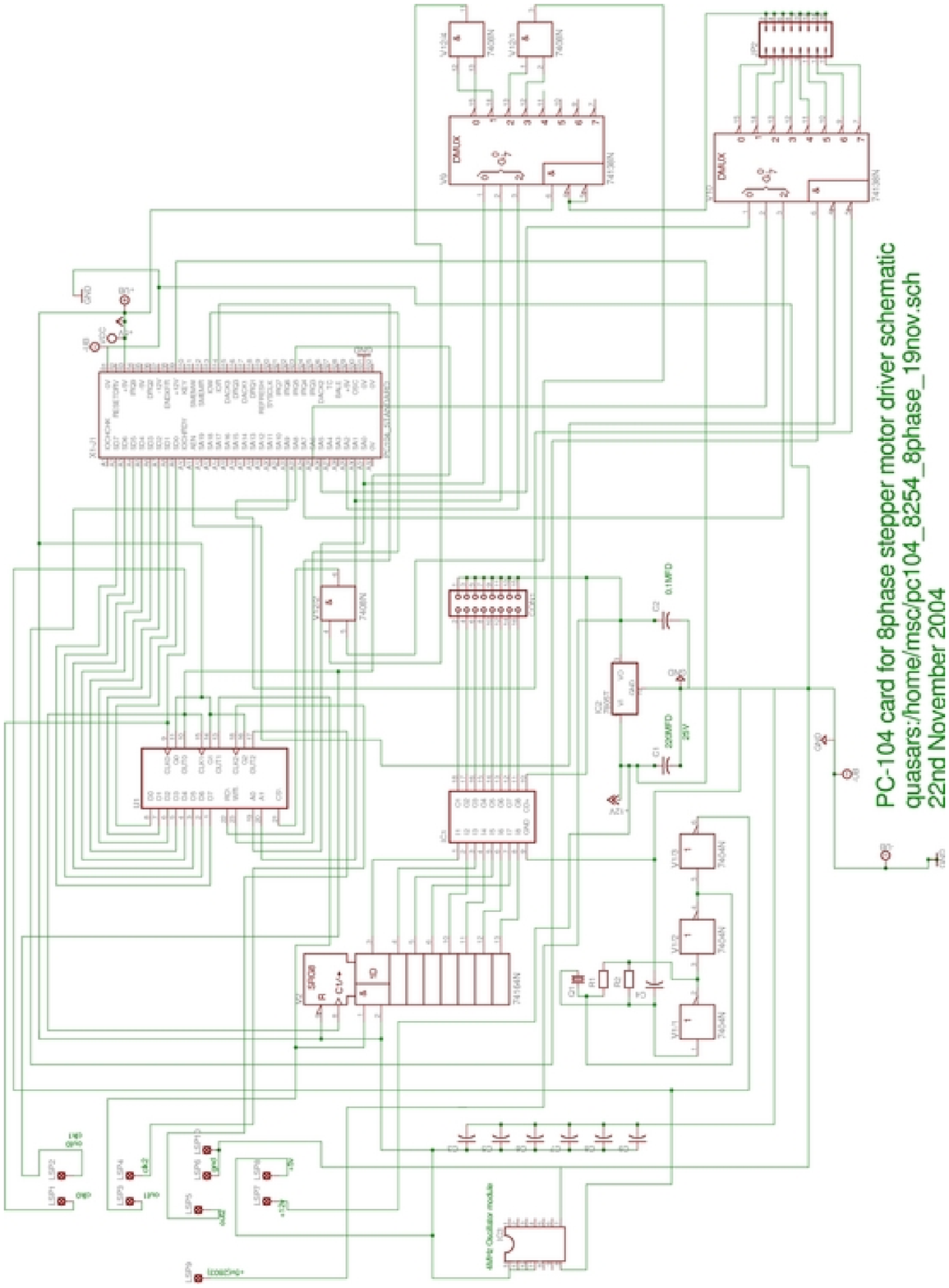}
\caption[Schematic of 8-phase stepper driver board]{\label{sch8phase}
Schematic of the 8-phase stepper driver PC-104 board}
\end{figure*}

\begin{figure*}[hb] 
\centering
\includegraphics[height=0.7\textheight]{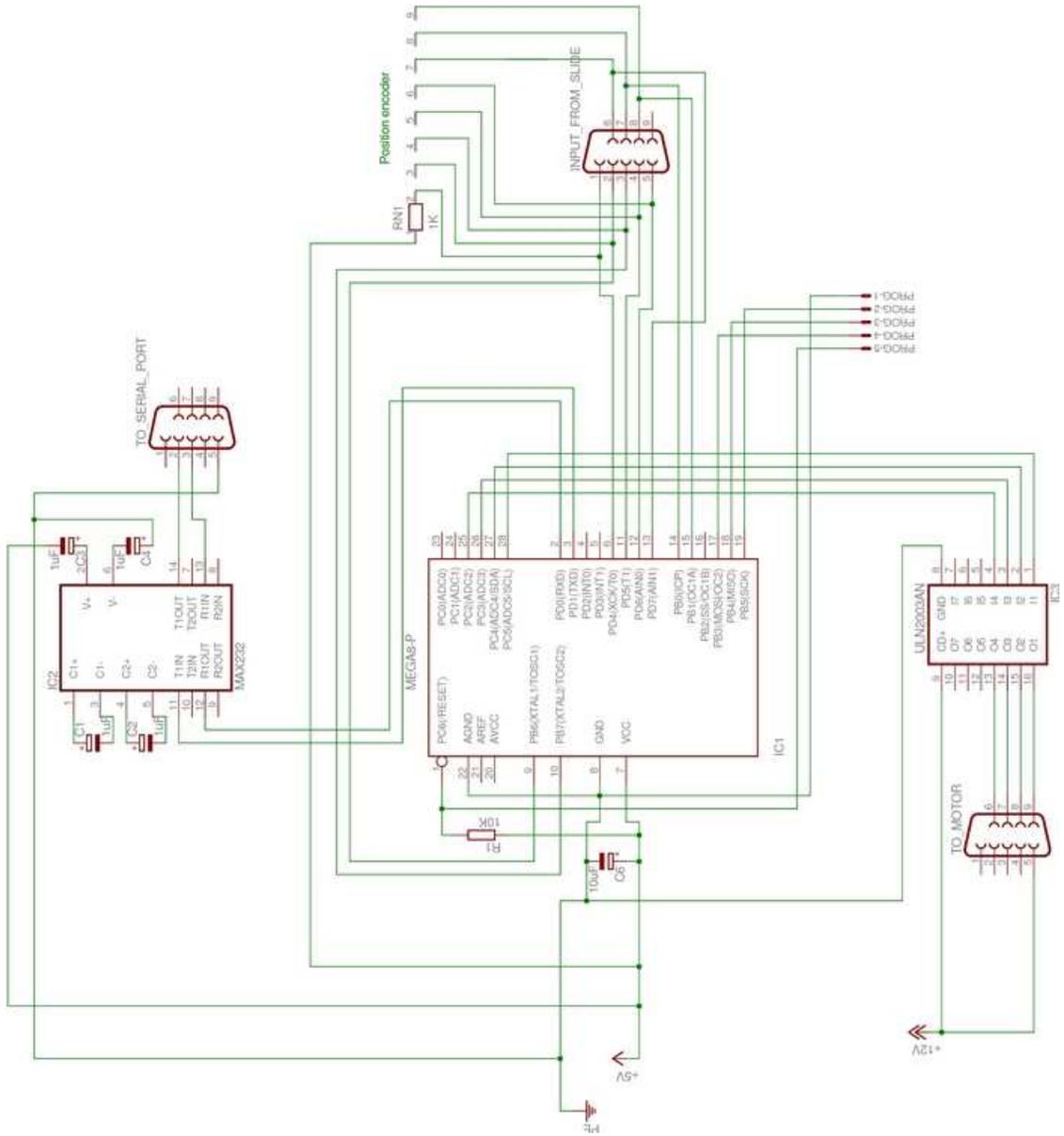}
\caption{\label{avrsch}
Schematic of the AVR micro-controller stepper driver board}
\end{figure*}
\end{twocolumn}
\end{document}